\pdfoutput=1
\def\year{2021}\relax
\documentclass[letterpaper]{article} % DO NOT CHANGE THIS
\usepackage{aaai21}  % DO NOT CHANGE THIS
\usepackage{times}  % DO NOT CHANGE THIS
\usepackage{helvet} % DO NOT CHANGE THIS
\usepackage{courier}  % DO NOT CHANGE THIS
\usepackage[hyphens]{url}  % DO NOT CHANGE THIS
\usepackage{graphicx} % DO NOT CHANGE THIS
\usepackage{natbib}  % DO NOT CHANGE THIS AND DO NOT ADD ANY OPTIONS TO IT
\usepackage{caption} % DO NOT CHANGE THIS AND DO NOT ADD ANY OPTIONS TO IT
\usepackage{outlines}

\usepackage{hyperref}
\usepackage{breakurl}
\usepackage[nobiblatex]{xurl}
\PassOptionsToPackage{hyphens}{url}

\title{The Limits of Global Inclusion in AI Development} % title proposal
\author{
    Alan Chan \thanks{Author order is alphabetical by last name. All authors contributed equally to this work.} \textsuperscript{\rm 1},
    Chinasa T. Okolo $^*$ \textsuperscript{\rm 2},
    Zachary Terner $^*$ \textsuperscript{\rm 3},
    Angelina Wang $^*$ \textsuperscript{\rm 4} 
    \\
}
\affiliations{
    \textsuperscript{\rm 1}Mila, Université de Montréal\quad
    \textsuperscript{\rm 2}Cornell University\quad\\
    \textsuperscript{\rm 3}National Institute of Statistical Sciences\quad
    \textsuperscript{\rm 4}Princeton University\quad
    
    alan.chan@mila.quebec, chinasa@cs.cornell.edu, zterner@niss.org, angelina.wang@princeton.edu
    
}
\begin{document}

\maketitle

\begin{abstract}
Those best-positioned to profit from the proliferation of artificial intelligence (AI) systems are those with the most economic power. Extant global inequality has motivated Western institutions to involve more diverse groups in the development and application of AI systems, including hiring foreign labour and establishing extra-national data centres and laboratories. However, given both the propensity of wealth to abet its own accumulation and the lack of contextual knowledge in top-down AI solutions, we argue that more focus should be placed on the redistribution of power, rather than just on including underrepresented groups. Unless more is done to ensure that opportunities to lead AI development are distributed justly, the future may hold only AI systems which are unsuited to their conditions of application, and exacerbate inequality. 
\end{abstract}

\section{Introduction}
% claim: inclusion itself isn’t enough to solve X problems in the long-term, if there is unequal distribution of roles in the AI ecosystem
The arm of global inequality is long, rendering itself visible especially in the development of artificial intelligence (AI). In an analysis of publications at two major machine learning conference venues, NeurIPS 2020 and ICML 2020, \citet{chuvpilo_ai_2020} found that of the top 10 countries in terms of publication index,
% \footnote{Calculated by treating a publication as a unit of 1, and splitting up the unit equally by authorship. See \citet{chuvpilo_ai_2020} for more details. }
none were located in Latin America, Africa, or Southeast Asia. Vietnam, the highest placing country of these groups, comes in 27th place. Of the top 10 institutions by publication index, eight out of 10 were based in the United States, including American tech giants like Google, Microsoft, and Facebook. Indeed, the full lists of the top 100 universities and top 100 companies by publication index include no companies or universities based in Africa or Latin America. Although conference publications are just one metric, they remain the predominant medium in which progress in AI is disseminated, and as such serve to be a signal of who is generating research. 

These statistics are unsurprising. The predominance of the United States in these rankings is consistent with its economic and cultural dominance, just as the appearance of China with the second highest index is a marker of its growing might. Also comprehensible is the relative absence of countries in the Global South, given the exploitation and underdevelopment of these regions by European colonial powers \citep{frank_capitalism_1967,rodney_how_1972,jarosz_human_2003,bruhn_good_2012}. 

Current global inequality in AI development involves both a concentration of profits and a danger of ignoring the contexts to which AI is applied. As AI systems become increasingly integrated into society, those responsible for developing and implementing such systems stand to profit to a large extent. If these players are predominantly located outside of the Global South, a disproportionate share of economic benefit will fall also outside of this region, exacerbating extant inequality. Furthermore, the ethical application of AI systems requires knowledge of the contexts in which they are to be applied. As recent work \citep{grush_google_2015,de_la_garza_states_2020,noauthor_abolish_2020,beede_human-centered_2020,sambasivan_non-portability_2020} has highlighted, work that lacks this contextual knowledge can fail to help the targeted individuals, and can even harm them (e.g., misdiagnoses in medical applications). 

Whether explicitly in response to these problems or not, calls have been made for broader inclusion in the development of AI \citep{asemota,lee_webuildai_2019}. At the same time, some have acknowledged the limitations of inclusion. \citet{sloane_participation_2020} describes and argues against participation-washing, whereby the mere fact that somebody has participated in a project lends it moral legitimacy. In this work, we focus upon the implications of participation for global inequality, focusing particularly on the limitations in which inclusion in AI development is practised in the Global South. We look specifically at how this plays out in the domains of datasets and research labs, and conclude with a discussion of opportunities for ameliorating the power imbalance in AI development.

\section{Datasets}
Given the centrality of large amounts of data in today's machine learning systems, there would appear to be substantial opportunity for inclusion in data collection and labeling processes. While there are benefits to more diverse participation in data-gathering pipelines (that is, processes involved in the collection, labeling, and other processing of data for use in machine-learning systems), we will highlight how this approach does not go far enough in addressing global inequality in AI development. 

% AC: commenting out for now for the proposed paragraph above
% As we consider ways in which to go about the inclusion of people from the Global South, the most basic, surface-level form of inclusion one could imagine is data labeling. We will discuss what kinds of problems this can alleviate, what the form of data labeling currently looks like, the barriers to participation, and also, the deep problems with this form of inclusion.

Data collection itself is a practice fraught with problems of inclusion and representation. Two large, publicly available image datasets, ImageNet~\citep{imagenet_cvpr09, imagenet} and OpenImages~\citep{openimages}, are US- and Eurocentric~\citep{shankar17}. \citet{shankar17} further argues that models trained on these datasets perform worse on images from the Global South. For example, images of grooms are classified with lower accuracy when they come from Ethiopia and Pakistan, compared to images of grooms from the United States. Along this vein, \citet{devries2019global} shows that images of the same word, like ``wedding" or ``spices", look very different when queried in different languages, as they are presented distinctly in different cultures. Thus, publicly available object recognition systems fail to correctly classify many of these objects when they come from the Global South. A representative dataset is crucial to allowing models to learn how certain objects and concepts are represented in different cultures. 

Since many deep learning techniques require large amounts of data to train their models, the importance of data labeling has grown. The data collection and labeling market is expected to grow to \$6.5 billion USD by 2027 \citep{grandviewresearch}, while \citet{cognilytica} estimates that over 80\% of the machine learning development process consists of data preparation tasks (collection, cleaning, and labeling). Large tech companies such as Uber and Alphabet rely heavily these services, with some paying millions of dollars monthly \citep{synced}. 

At the same time, data labeling is a time-consuming, repetitive process. Its importance in machine-learning research and development has led to the crowdsourcing of this work, whereby anonymous individuals are remunerated for completing this work. A major venue for crowdsourcing work is Amazon Mechanical Turk; according to \citet{difallah18mt}, less than 2\% of Mechanical Turk workers come from the Global South (a vast majority come from the USA and India). Other notable companies in this domain, Samasource, Scale AI, and Mighty AI also operate in the United States, but they crowdsource workers from around the world, primarily relying on low-wage workers from sub-Saharan Africa and Southeast Asia \cite{murgia2019ai}. This leads to a significant disparity between the millions in profits earned by data labeling companies and worker earnings; for example, workers at Samasource earn around \$8 USD a day \citep{lee2018big} while the company made \$19 million in 2019 \cite{samasourceCauseIQ}. While \citet{lee2018big} notes that \$8 USD may well be a living wage in certain areas, the massive profit disparity remains despite the importance of these workers to the core businesses of these companies. Additionally, many of these workers are contributing to AI systems that are likely to be biased against underrepresented populations in the locales they are deployed in~\cite{buolamwini2018gender, obermeyer2019dissecting} and may not be directly benefiting their local communities. While data labeling is not as physically intensive as traditional factory labor, workers report the pace and volume of their tasks as "mentally exhausting" and "monotonous" due to the strict requirements needed for labeling images, videos, and audio to client specifications \cite{bbcghostwork, weflowskill}. In the Global South recently, local companies have begun to proliferate, like Fastagger in Kenya, Sebenz.ai in South Africa, and Supahands in Malaysia. As AI development continues to scale, the expansion of these companies opens the door for low-skilled laborers to enter the workforce but also presents a chance for exploitation to continue to occur. %With recent initiatives in responsible and trustworthy AI by IBM, Microsoft, DeepMind, and more \cite{msftrespAI, IBMtrustAI, DeepMindsafeAI}, we hope these methodologies will apply not only to the development of machine learning algorithms but also extend to the workers on whom these models depend.
% some relevant papers specific to this section that might be helpful:
% https://www.pewresearch.org/internet/2016/07/11/turkers-in-this-canvassing-young-well-educated-and-frequent-users/

% https://www.ipeirotis.com/wp-content/uploads/2017/12/wsdmf074-difallahA.pdf (mechanical turk demographics)

\subsubsection{Barriers to Participation}

% Section 1C: Barriers to participating
There are barriers that exist to participating in data labeling. The most obvious is that a computing device and stable internet access are required for access to these data labeling platforms. These goods are highly correlated with socioeconomic status and geographic locations, thus serving as a barrier to participation for many \citet{harris2017socioeconomic}. A reliable internet connection is necessary for finding tasks to complete, completing those tasks, and accessing the remuneration for those tasks. Further, those in the Global South pay higher prices for Internet access compared to their counterparts in the Global North (i.e. Western countries) \cite{nzekwe2019}. Another barrier is in the method of payment for data labeling services on some of these platforms. For example, Amazon Mechanical Turk, a widely used platform for finding data labelers, only allows payment to a U.S. Bank Account or in the form of an Amazon.com gift card~\citep{amtfaq}. These methods of payment restrict may not be what is desired by a worker, and can serve as a deterrent to work for this platform.

\subsubsection{Problems with Participation}
% After having discussed the benefits that incorporating data labeling as one slice of inclusion can have, as well as some of the barriers to participation it has, 
% We finish this section by discussing many of the problematic issues that come with data labeling. These issues will need to be altered before data labeling constitutes a valid portion of the significant process of inclusion.
% At a cursory glance, having labelers who represent a diversity of backgrounds might appear largely beneficial, as it would allow for objects that might not be recognized and labeled appropriately (e.g., ``wedding") by one group of people to be done so by another. Additionally, data labelers are prone to bringing their own stereotypes and biases to the task at hand. Diversifying the labeler population would at least help dilute the pool of shared biases that could propagate into the dataset. For example, it has been shown that MSCOCO~\citep{lin14coco}, a commonly-used object detection and image captioning dataset, contains strong gender biases in the image captions~\citep{hendricks2018snowboard, bhargava2019caption}. 
% If the population of dataset labelers comprised people more aware of the problems with gender stereotypes, or even people with very different gender stereotypes, perhaps the biases in the captions might not manifest with that level of prevalence.

Although global inclusion in the data pipeline can be beneficial, it is no panacea for global inequality in AI development, and in fact, can even be detrimental if not approached with care. The development of AI is highly concentrated in countries in the Global North for a variety of reasons, such as an abundance of capital, well-funded research institutions, and technical infrastructure. The existence of these advantageous conditions is inextricable from the history of colonial exploitation of the Global South, whereby European states plundered labour and capital for the benefit of the metropoles, to the detriment of the colonized \citep{frank_capitalism_1967,rodney_how_1972}. A key justification for this exploitation was white supremacy: the colonized, as ``uncivilized", were most fit to perform physically excruciating labour, at wages lower than those paid to Europeans. As such, colonized peoples were for the most part prevented from engaging in the more lucrative businesses of insurance, banking, industry, and trading \citep{rodney_how_1972}. Although the labour and natural capital of colonized nations were indispensable to European economic projects, European institutions and individuals captured the vast majority this wealth. 

It is instructive to view inclusion in the data pipeline as a continuation of this exploitative history. With respect to data collection, current practices can neglect consent and poorly represent areas of the Global South. Image datasets are often collected without consent from the people involved, even in pornographic contexts~\citep{Prabhu20, paullada20data}, while others (e.g., companies, end-users) benefit from their use.~\citet{jo20archives} suggests drawing from the long tradition or archives when collecting data because this is a discipline that has already been thinking about challenges like consent and privacy. Indeed, beyond a possible honorarium for participation in the data collection process, no large-scale, successful schema currently exists for compensating users for the initial and continued use of their data in machine-learning systems, although some efforts are currently underway \citep{kelly_andrew_2020}. However, the issue of compensation elides the question of whether such large-scale data collection should occur in the first place. Indeed, the process of data collection can contribute to an ``othering'' of the subject and cement inaccurate or harmful beliefs. Even if data come from somewhere in the Global South, they are often from the perspective of an outsider \citep{wang20revise}. That the outsider may not understand the context or may have an agenda counter to the interest of the subject is reflected in the data captured, as has been extensively studied in the case of photography \citep{ranger_colonialism_2001,batziou_framing_2011,thompson_otherness_2016}. Ignorance of context can cause harm, as \citet{sambasivan_non-portability_2020} discusses in the case of fair ML in India, where distortions in the data (e.g., a given sample corresponds to multiple individuals because of shared device usage) distort the meaning of fairness definitions that were formulated in Western contexts.
Furthermore, the history of phrenology reveals the role that the measurement and classification of colonial subjects had in justifying domination \citep{bank_native_1996,poskett_django_2013}.~\citet{denton20dataset} points out the need to interrogate more deeply the norms and values behind the creation of datasets, as they are often extractive processes that benefit only the dataset collector and users.

As another significant part of the data collection pipeline, data labeling is an extremely low-paying job involving rote, repetitive tasks that offer no room for upward mobility. Individuals may not require many technical skills to label data, but they do not develop any meaningful technical skills either. The anonymity of platforms like Amazon's Mechanical Turk inhibit the formation of social relationships between the labeler and the client that could otherwise have led to further educational opportunities or better remuneration. Although data is central to the AI systems of today, data labelers receive only a disproportionately tiny portion of the profits of building these systems. In parallel with colonial projects of resource extraction, data labeling as extraction of meaning from data is no way out of a cycle of colonial dependence. 
% Exporting these kinds of jobs follows in the long history of colonialism~\citep{mohamed2020decolonial}, with the groups on the receiving end of the labor showing great gains in the form of strong artificial intelligence models, but the groups on the giving end of the labor receiving few of the benefits of their work.

The people doing the work of data labeling have been termed "ghost-workers"~\citep{gray2019ghostwork}. The labour of these unseen workers generates massive profits that others capture. While our following discussion provides US statistics because those are the ones most readily available, it is easy to imagine similar or worse labour situations in the Global South. ImageNet~\citep{imagenet_cvpr09, imagenet}--a benchmark dataset essential to recent progress in computer vision--would have not been possible without the work of data labelers~\citep{gershgorn2017imagenet}. However, the workers themselves made only around a median of \$2/hour USD, with only 4\% making more than the US federal minimum wage of \$7.25/hour~\citep{hara18amt}, itself a far cry from a living wage. The study attributed much of this low-wage structure to the time spent on activities that were not compensated, such as finding tasks or working on tasks that are ultimately rejected. This leads into another major problem of the power dynamics on a platform like Amazon Mechanical Turk, where all of the power is given to the requester of the task. Requesters have the power to set any price they want (as low as \$.01), reject the completed work of a worker, and misleadingly claim their task will take a length of time much shorter than what it would actually take~\citep{semuels2018internet}. In the US, workers in this business are considered independent contractors rather than employees, so protections guaranteed by the Fair Labor Standards Act do not apply. A same lack of protections can be seen for data labelers in the Global South~\citep{kaye19dataannot}. This power imbalance emphasizes the need for labor protection. 

%Online gig work~\citep{graham2018gig} is developing at a rapid rate, and it is vital that the contributions of these ghost workers be valued. Currently, data labelers are often wholly detached from the rest of the ML pipeline, with workers oftentimes not knowing how their labor will be used nor for what purpose \citep{graham2018planteary}. Little sense of fulfillment comes from menial tasks, and by exploiting these workers solely for their produced knowledge without bringing them into the fold of the product that they are helping to create, a deep chasm exists between workers and the downstream product~\citep{rogstadius2011motivation}. Thus, in addition to policy that improves work conditions and wages for data labelers, workers should be provided with education opportunities that allow them to contribute to the models they are building in ways beyond labeling. They should be both incorporated into the process in ways which support the idea that their work is valued, and provided with opportunities for career advancement~\citep{gray2019ghostwork}. Valuing data labeling work in this economic form, of increasing wages and allowing professional mobility, backs up an otherwise empty statements of valuation. 

\section{Research Labs}

Establishing research labs has been essential for major tech companies to advance the development of their respective technologies while providing valuable contributions to the field of computer science \cite{natureIRL}. In the United States, General Electric (GE) Research Laboratory is widely accepted as the first industrial research lab, providing early technological achievements to GE and establishing them as a leader in industrial innovation \cite{center7general}. As the ascendance of artificial intelligence becomes more important to the bottom lines of many large tech companies, industrial research labs have spun out that solely focus on artificial intelligence and its respective applications. Companies from Google to Amazon to Snapchat have doubled down in this field and opened up labs leveraging artificial intelligence for web search, language processing, video recognition, voice applications, and much more. As AI becomes increasingly integrated into the livelihoods of consumers around the world, tech companies have recognized the importance of democratizing AI development and moving it outside the bounds of the Global North. Of five notable tech companies developing AI solutions (Google, Microsoft, IBM, Facebook, and Amazon), Google, Microsoft, and IBM have research labs in the Global South and all have either development centers, customer support centers, or data centers within these regions. Despite their presence throughout the Global South, AI research centers tend to be concentrated in certain countries. Within Southeast Asia, the representation of lab locations is limited to India; in South America, representation is limited to Brazil. In sub-Saharan Africa we find a bit more spread in location with AI labs established in Accra, Ghana; Nairobi, Kenya; and Johannesburg, South Africa.

\subsubsection{Barriers to Participation}
For a company to choose to establish an AI research center, the company must believe this initiative to be in its financial interest. Unfortunately, several barriers exist. The necessity of generating reliable returns for shareholders precludes ventures that appear too risky, especially for smaller companies. The perception of risk can take a variety of forms and possibly be influenced by stereotypes to differing extents. Two such factors are political/economic instability or a relatively lower proportion of tertiary formal education in the local population, which can be traced to the history of colonial exploitation and underdevelopment \citep{rodney_how_1972,jarosz_human_2003, bruhn_good_2012}, whereby European colonial powers extracted labour, natural resources, and economic surplus from colonies, while at the same time subordinating their economic development to that of the metropoles. It is hard to imagine the establishment of a top-tier research university --- with the attendant technical training afforded to the local populace --- in regions repeatedly denuded of wealth. 

\subsubsection{Problems with Participation}
While the opening of data centers and AI research labs in the Global South appears beneficial for the local workforce, these positions may require technical expertise which the local population might not have. This would instead introduce opportunities for displacement by those from the Global North who have had more access to specialized training needed to develop, maintain, and deploy AI systems. Given the unequal distribution of AI development globally, it is common for AI researchers and practitioners to work and study in places outside of their home countries (i.e., outside of the Global South). For example, the current director of Google AI Accra, originally from Senegal, was recruited to Google from Facebook AI Research in Menlo Park, CA~\cite{adekanmbi_2018,asemota}. The director for Microsoft’s new lab in Nairobi, Kenya was recruited from Microsoft Research India; before that, she was a research scientist at Xerox in France~\cite{jacki, jackimsr}. While the directors of many research labs established in the Global South have experience working in related contexts, we find that local representation is sorely lacking at both the leadership and general workforce level. Grassroots AI education and training initiatives by communities such as Deep Learning Indaba, Data Science Africa, and Khipu AI in Latin America aim to increase local AI talent, but since these initiatives are less than five years old, it is hard to measure their current impact on improving the pipeline of AI researchers and machine learning engineers. However, with the progress made by these organizations publishing novel research at premier AI conferences, hosting conferences of their own, and much more, the path to inclusive representation in the global AI workforce is strengthening.

Although several tech companies have established research facilities across the world and in the Global South, these efforts remain insufficient at addressing long-term problems in the AI ecosystem. A recent report from Georgetown University's Center for Security and Emerging Technologies (CSET) describes the establishment of AI labs by US companies, namely Facebook, Google, IBM, and Microsoft, abroad~\cite{cset}. The report notes that while 68\% of the 62 AI labs are located outside of the United States, 68\% of the staff are located within the United States. Therefore, the international offices remain half as populated on average relative to the domestic locations. Additionally, none of these offices are located in South America and only four are in Africa. To advance equity within AI and improve inclusion efforts, it is imperative that companies not only establish locations in underrepresented regions, but hire employees and include voices from those regions in a proportionate manner. 

The CSET report also notes that AI labs form abroad generally in one of three ways: through the acquisition of startups; by establishing partnerships with local universities or institutions; and by relocating internal staff or hiring new staff in these locations~\cite{cset}. The first two of these methods may favor locations with an already-established technological or AI presence, as many AI startups are founded in locations where a financial and technological support system exists for them. Similarly, the universities with whom tech companies choose to partner are often already leaders in the space, as evidenced by Facebook's partnership with Carnegie Mellon professors and MIT's partnerships with both IBM and Microsoft. The general strategy of partnering with existing institutions and of acquiring startups has the potential to reinforce existing inequities by investing in locations with already thriving tech ecosystems. One notable exception to this is Google's investment into infrastructure, skills training, and startups in Ghana~\cite{asemota}. Long-term investment and planning in the Global South can form the stepping stones for broadening AI to include underrepresented and marginalized communities.  

Even with long-term investment into regions in the Global South, the question remains of whether local residents are provided opportunities to join management and contribute to important strategic decisions. Several organizations have emphasized the need for AI development within a country to happen at the grassroots level, so that those implementing AI as a solution understand the context of the problem being solved~\cite{mbayo_2020, gul_2019}. The necessity of indigenous decision-making is just as important in negotiating the values that AI technologies are to instantiate, such as through AI ethics declarations that are at the moment heavily Western-based \citep{jobin_global_2019}. Although this is critical not only to the success of individual AI solutions but also to equitable participation within the field at large, more can and should be done. True inclusion necessitates that underrepresented voices can be found in all ranks of a company's hierarchy, including in positions of upper management. Tech companies which are establishing a footprint in these regions are uniquely positioned to offer this opportunity to natives of the region. Taking advantage of this ability will be critical to ensuring that the benefits of AI apply not only to technical problems that arise in the Global South, but to socioeconomic inequalities which persist around the world.

\section{Opportunities}
In the face of global inequality in AI development, there are a few promising opportunities.

\subsubsection{Affinity Groups}
While AI and technology in general has long excluded marginalized populations, the emergence of grassroots efforts by organizations to ensure that indigenous communities are actively involved as stakeholders of AI has recently been strong. Black in AI, a nonprofit organization with worldwide membership, was founded to increase the global representation of Black-identifying students, researchers, and practitioners in the field of AI, and has made significant improvements in increasing the number of Black scholars attending and publishing in NeurIPS and other premier AI conferences~\cite{earl_2020,BlackInAI}. Inclusion in AI is extremely sparse in higher education and recent efforts by Black in AI have focused on instituting programming to support members in graduate programs and in their postgraduate careers. Other efforts such as Khipu AI, based in Latin America, have been established to provide a venue to train aspiring AI researchers in advanced machine learning topics, foster collaborations, and actively participate in how AI is being used to benefit Latin America. Other communities based on the African continent such as Data Science Africa and Deep Learning Indaba have expanded their efforts, establishing conferences, workshops, and dissertation awards, and developing curricula for the broader African AI community. These communities are clear about their respective missions and the focus of collaboration. Notably, Masakhane, a grassroots organization focusing on improving the representation of African languages in the field of natural language processing shares the sentiment expressed in this paper on how AI research should be approached:

\begin{quote}
Masakhane are not just annotators or translators. We are researchers. We can likely connect you with annotators or translators but we do not support shallow engagement of Africans as only data generators or consumers \citep{noauthor_masakhane_nodate}.
\end{quote}

As these initiatives grow across the Global South, we hope large organizations and technology companies partner with and adopt the values of these respective initiatives to ensure AI developments are truly representative of the global populace.

\subsubsection{Research Participation}
One key component of AI inclusion efforts should be to elevate the involvement and participation of those historically excluded from technological development. Many startups and several governments across the Global South are creating opportunities for local communities to participate in the development and implementation of AI programs~\cite{mbayo_2020, gul_2019, galperin_alarcon}. In situations where the central involvement has been data labeling, strides should be taken to add model development roles to the opportunity catalog there. Currently, data labelers are often wholly detached from the rest of the ML pipeline, with workers oftentimes not knowing how their labor will be used nor for what purpose \citep{graham2018planteary}. Little sense of fulfillment comes from menial tasks, and by exploiting these workers solely for their produced knowledge without bringing them into the fold of the product that they are helping to create, a deep chasm exists between workers and the downstream product~\citep{rogstadius2011motivation}. Thus, in addition to policy that improves work conditions and wages for data labelers, workers should be provided with education opportunities that allow them to contribute to the models they are building in ways beyond labeling~\citep{gray2019ghostwork}. Similarly, where participation in the form of model development is the norm, employers should seek to involve local residents in the ranks of management and in the process of strategic decision-making. The advancement of an equitable AI workforce and ecosystem requires that those in positions of data collection and training be afforded opportunities to lead their organizations. Including these voices in positions of power has the added benefit of ensuring the future hiring and promotion of local community members.

\subsubsection{AI as Development}
% comparison to ISI
% ISI is a strategy for relatively underdeveloped countries to industrialize quickly, in order to realize the productivity gains of industrialization
% AI development is sort of analagous, esp since the research/engineering jobs pay a higher wage than data labeling or lower parts on the AI usage hierarchy
% ISI was hard to carry out though, and depends on key factors: Mendes et al. 2014, Aryeetey and Moyo 2012
% requires massive investment for education, but perhaps not as much as you would think (i.e., don't have to have google-level resources to do good research or to implement AI systems to help people)

The massive inequalities in the development of AI can appear daunting. Will it ever be possible to close the gap? Similar concerns arise in the broader study of economic development, from which one can draw lessons. 

Despite the large developmental gap between the Global North and the Global South, the latter part of the 20th century saw some countries bridge it. For example, while the GDP per capita of South Korea was far lower than that of the USA in the 1960s, by 2000 the gap had considerably narrowed, especially in comparison to world GDP per capita over the same time period. \footnote{\url{https://ourworldindata.org/grapher/average-real-gdp-per-capita-across-countries-and-regions?time=1869..2016&country=KOR~USA~OWID_WRL}} Much work \citep{chang_bad_2009,lin_flying_2011,aryeetey_industrialisation_2012,mendes_industrialization_2014} has linked the relative economic success of South Korea to the policy of import substitution industrialization (ISI), whereby a country attempts to replace foreign imports with domestic production in an attempt to build high-productivity industries (e.g., electronics), rather than rely on exports of low-productivity industries (e.g., agriculture). The idea is that once the so-called ``infant industries" have developed enough, they will be able to compete in international markets without government support. The execution of ISI involves protectionist trade policies, subsidies for targeted industries, and sufficient investment in education and infrastructure. While ISI can be incredibly successful, as in the cases of Samsung and POSCO from South Korea \citep{chang_bad_2009}, its execution relies on sufficient agricultural input and human capital, careful management of foreign reserves, and state capacity for coordination with private partners \citep{aryeetey_industrialisation_2012,mendes_industrialization_2014}. In the absence of these factors, ISI can fail and the country can even go through de-industrialization. 

We suggest viewing AI development as a path forward for economic development, in light of the lessons learned from ISI policies. Rather than rely upon foreign construction of AI systems for domestic application, where any returns from these systems are not reinvested domestically, we encourage the formation of domestic AI development activity. This development activity should not be focused on low-productivity activities, such as data-labeling, but instead on high-productivity activities like model development/deployment and research. An AI-focused ISI policy could include state-led investments into AI-related education and infrastructure, funding for private bodies to engage in domestic AI development, and limitations on the extent to which foreign companies may be involved in or profit from domestic AI activities. While it remains essential, as it was in historical ISI policies, to work with and assimilate technology and expertise from foreign companies, it is imperative that domestic expertise be developed in tandem to shape the future of AI development and reap its large profits.  

This is by no means an easy task, and an AI-focused ISI policy encounters many of the same difficulties as historical ISI policies, such as the necessity of bringing in expertise and technology, and in ensuring that sufficient education and infrastructure (e.g., internet access) exist. It will likely encounter many new difficulties that are unique to AI development as well. Even in the absence of centralized state coordination, however, recent initiatives like Deep Learning Indaba and Khipu have promoted the importance of indigenous AI development and have advanced education in AI.

\section{Conclusion}
As the development of artificial intelligence continues to progress across the world, the exclusion of those from communities most likely to bear the brunt of algorithmic inequity only stands to worsen. We address this question by exploring the challenges and benefits of increasing broader inclusion in the field of AI. We examine the limits of current AI inclusion methods, problems of participation regarding AI labs situated in the Global South from major tech companies, and discuss opportunities for AI to accelerate development within disadvantaged regions.

We hope the actions we propose can help to begin the movement of communities in the Global South from being just beneficiaries or subjects of AI systems to being active, engaged participants. Having true agency over the AI systems integrated into the livelihoods of communities in the Global South will maximize the impact of these systems and lead the way for global inclusion of AI.

As a limitation of our work, it is important to acknowledge we are currently all located at, and have been educated at, North American institutions. Our positions in these institutions thus limit our perspective, and we respect the considerations we may have missed and the voices we have not heard in the course of writing this work.

\bibliography{refs.bib}

\begin{thebibliography}{67}
\providecommand{\natexlab}[1]{#1}
\providecommand{\url}[1]{\texttt{#1}}
\providecommand{\urlprefix}{URL }
\expandafter\ifx\csname urlstyle\endcsname\relax
  \providecommand{\doi}[1]{doi:\discretionary{}{}{}#1}\else
  \providecommand{\doi}{doi:\discretionary{}{}{}\begingroup
  \urlstyle{rm}\Url}\fi

\bibitem[{Adekanmbi(2018)}]{adekanmbi_2018}
Adekanmbi, B. 2018.
\newblock 10 inspiring Facts about Moustapha Cisse, Google AI Ghana Pioneer
  Lead.
\newblock
  \urlprefix\url{https://www.datasciencenigeria.org/10-inspiring-facts-moustapha-cisse-google-ai-ghana-pioneer-lead/}.

\bibitem[{Amazon(2020)}]{amtfaq}
Amazon. 2020.
\newblock FAQs.
\newblock \urlprefix\url{https://www.mturk.com/worker/help}.

\bibitem[{Aryeetey and Moyo(2012)}]{aryeetey_industrialisation_2012}
Aryeetey, E.; and Moyo, N. 2012.
\newblock Industrialisation for {Structural} {Transformation} in {Africa}:
  {Appropriate} {Roles} for the {State}.
\newblock \emph{Journal of African Economies} 21(suppl\_2): ii85.
\newblock
  \urlprefix\url{https://econpapers.repec.org/article/oupjafrec/v_3a21_3ay_3a2012_3ai_3asuppl_5f2_3ap_3a-ii85.htm}.
\newblock Publisher: Centre for the Study of African Economies (CSAE).

\bibitem[{Asemota(2018)}]{asemota}
Asemota, V. 2018.
\newblock 'Ghana is the future of Africa': Why Google built an AI lab in Accra.
\newblock
  \urlprefix\url{https://edition.cnn.com/2018/07/14/africa/google-ghana-ai/}.

\bibitem[{Bank(1996)}]{bank_native_1996}
Bank, A. 1996.
\newblock Of '{Native} {Skulls}' and '{Noble} {Caucasians}': {Phrenology} in
  {Colonial} {South} {Africa}.
\newblock \emph{Journal of Southern African Studies} 22(3): 387--403.
\newblock ISSN 0305-7070.
\newblock \urlprefix\url{http://www.jstor.org/stable/2637310}.
\newblock Publisher: [Taylor \& Francis, Ltd., Journal of Southern African
  Studies].

\bibitem[{Batziou(2011)}]{batziou_framing_2011}
Batziou, A. 2011.
\newblock Framing ‘otherness’ in press photographs: {The} case of
  immigrants in {Greece} and {Spain}.
\newblock \emph{Journal of Media Practice} 12(1): 41--60.
\newblock ISSN 1468-2753.
\newblock \doi{10.1386/jmpr.12.1.41_1}.
\newblock \urlprefix\url{https://doi.org/10.1386/jmpr.12.1.41_1}.
\newblock Publisher: Routledge \_eprint:
  https://doi.org/10.1386/jmpr.12.1.41\_1.

\bibitem[{Beede et~al.(2020)Beede, Baylor, Hersch, Iurchenko, Wilcox,
  Ruamviboonsuk, and Vardoulakis}]{beede_human-centered_2020}
Beede, E.; Baylor, E.; Hersch, F.; Iurchenko, A.; Wilcox, L.; Ruamviboonsuk,
  P.; and Vardoulakis, L.~M. 2020.
\newblock A {Human}-{Centered} {Evaluation} of a {Deep} {Learning} {System}
  {Deployed} in {Clinics} for the {Detection} of {Diabetic} {Retinopathy}.
\newblock In \emph{Proceedings of the 2020 {CHI} {Conference} on {Human}
  {Factors} in {Computing} {Systems}}, {CHI} '20, 1--12. New York, NY, USA:
  Association for Computing Machinery.
\newblock ISBN 978-1-4503-6708-0.
\newblock \doi{10.1145/3313831.3376718}.
\newblock \urlprefix\url{http://doi.org/10.1145/3313831.3376718}.

\bibitem[{Bruhn and Gallego(2012)}]{bruhn_good_2012}
Bruhn, M.; and Gallego, F.~A. 2012.
\newblock Good, {Bad}, and {Ugly} {Colonial} {Activities}: {Do} {They} {Matter}
  for {Economic} {Development}?
\newblock \emph{The Review of Economics and Statistics} 94(2): 433--461.
\newblock
  \urlprefix\url{https://ideas.repec.org/a/tpr/restat/v94y2012i2p433-461.html}.
\newblock Publisher: MIT Press.

\bibitem[{Buolamwini and Gebru(2018)}]{buolamwini2018gender}
Buolamwini, J.; and Gebru, T. 2018.
\newblock Gender shades: Intersectional accuracy disparities in commercial
  gender classification.
\newblock In \emph{Conference on fairness, accountability and transparency},
  77--91.

\bibitem[{Center(2011)}]{center7general}
Center, E.~T. 2011.
\newblock General Electric Research Lab.
\newblock \urlprefix\url{https://edisontechcenter.org/GEresearchLab.html}.

\bibitem[{Chang(2009)}]{chang_bad_2009}
Chang, H.-J. 2009.
\newblock \emph{Bad {Samaritans}: {The} {Myth} of {Free} {Trade} and the
  {Secret} {History} of {Capitalism}}.
\newblock New York, NY: Bloomsbury Press.
\newblock ISBN 978-1-59691-598-5.

\bibitem[{Chuvpilo(2020)}]{chuvpilo_ai_2020}
Chuvpilo, G. 2020.
\newblock {AI} {Research} {Rankings} 2020: {Can} the {United} {States} {Stay}
  {Ahead} of {China}?
\newblock
  \urlprefix\url{https://chuvpilo.medium.com/ai-research-rankings-2020-can-the-united-states-stay-ahead-of-china-61cf14b1216}.

\bibitem[{{Coalition for Critical Technology}(2020)}]{noauthor_abolish_2020}
{Coalition for Critical Technology}. 2020.
\newblock Abolish the \#{TechToPrisonPipeline}.
\newblock
  \urlprefix\url{https://medium.com/@CoalitionForCriticalTechnology/abolish-the-techtoprisonpipeline-9b5b14366b16}.

\bibitem[{Cognilytica(2019)}]{cognilytica}
Cognilytica. 2019.
\newblock Data Engineering, Preparation, and Labeling for AI.
\newblock
  \urlprefix\url{https://www.cognilytica.com/2019/03/06/report-data-engineering-preparation-and-labeling-for-ai-2019/}.

\bibitem[{Croce and Musa(2019)}]{weflowskill}
Croce, N.; and Musa, M. 2019.
\newblock The new assembly lines: Why AI needs low-skilled workers too.
\newblock
  \urlprefix\url{https://www.weforum.org/agenda/2019/08/ai-low-skilled-workers/}.

\bibitem[{De~La~Garza(2020)}]{de_la_garza_states_2020}
De~La~Garza, A. 2020.
\newblock States' {Automated} {Systems} {Are} {Trapping} {Citizens} in
  {Bureaucratic} {Nightmares} {With} {Their} {Lives} on the {Line}.
\newblock \urlprefix\url{https://time.com/5840609/algorithm-unemployment/}.

\bibitem[{Deng et~al.(2009)Deng, Dong, Socher, Li, Li, and
  Fei-Fei}]{imagenet_cvpr09}
Deng, J.; Dong, W.; Socher, R.; Li, L.-J.; Li, K.; and Fei-Fei, L. 2009.
\newblock {ImageNet: A Large-Scale Hierarchical Image Database}.
\newblock In \emph{CVPR}.

\bibitem[{Denton et~al.(2020)Denton, Hanna, Amironesei, Smart, Nicole, and
  Scheuerman}]{denton20dataset}
Denton, E.; Hanna, A.; Amironesei, R.; Smart, A.; Nicole, H.; and Scheuerman,
  M.~K. 2020.
\newblock Bringing the People Back In: Contesting Benchmark Machine Learning
  Datasets.
\newblock \emph{ICML Workshop on Participatory Approaches to Machine Learning}
  .

\bibitem[{DeVries et~al.(2019)DeVries, Misra, Wang, and van~der
  Maaten}]{devries2019global}
DeVries, T.; Misra, I.; Wang, C.; and van~der Maaten, L. 2019.
\newblock Does Object Recognition Work for Everyone?
\newblock \emph{Computer Vision and Pattern Recognition Workshop (CVPRW)} .

\bibitem[{Difallah, Filatova, and Ipeirotis(2018)}]{difallah18mt}
Difallah, D.; Filatova, E.; and Ipeirotis, P. 2018.
\newblock Demographics and Dynamics of Mechanical Turk Workers.
\newblock \emph{Proceedings of WSDM: The Eleventh ACM International Conference
  on Web Search and Data Mining} .

\bibitem[{Earl(2020)}]{earl_2020}
Earl, C.~C. 2020.
\newblock Notes from the Black In AI 2019 Workshop.
\newblock
  \urlprefix\url{https://charlesearl.blog/2020/01/08/notes-from-the-black-in-ai-2019-workshop/}.

\bibitem[{Frank(1967)}]{frank_capitalism_1967}
Frank, A.~G. 1967.
\newblock \emph{Capitalism and underdevelopment in {Latin} {America} :
  historical studies of {Chile} and {Brazil}.}
\newblock New York: Monthly Review Press.

\bibitem[{Galperin and Alarcon(2018)}]{galperin_alarcon}
Galperin, H.; and Alarcon, A. 2018.
\newblock The Future of Work in the Global South.

\bibitem[{Gent(2019)}]{bbcghostwork}
Gent, E. 2019.
\newblock The ‘ghost work’ powering tech magic.
\newblock
  \urlprefix\url{https://www.causeiq.com/organizations/samasource,262547062/}.

\bibitem[{Gershgorn(2017)}]{gershgorn2017imagenet}
Gershgorn, D. 2017.
\newblock The data that transformed AI research—and possibly the world.
\newblock \emph{Quartz} .

\bibitem[{Graham(2018)}]{graham2018planteary}
Graham, M. 2018.
\newblock The rise of the planetary labour market – and what it means for the
  future of work.
\newblock \emph{NS Tech} .

\bibitem[{{Grand View Research}(2020)}]{grandviewresearch}
{Grand View Research}. 2020.
\newblock Data Collection \& Labeling Market Size Worth $6.5$ Billion By 2027.
\newblock
  \urlprefix\url{https://www.grandviewresearch.com/press-release/global-data-collection-labeling-market}.

\bibitem[{Gray and Suri(2019)}]{gray2019ghostwork}
Gray, M.~L.; and Suri, S. 2019.
\newblock \emph{Ghost Work: How to Stop Silicon Valley from Building a New
  Global Underclass}.
\newblock Houghton Mifflin Harcourt.

\bibitem[{Grush(2015)}]{grush_google_2015}
Grush, L. 2015.
\newblock Google engineer apologizes after {Photos} app tags two black people
  as gorillas.
\newblock
  \urlprefix\url{https://www.theverge.com/2015/7/1/8880363/google-apologizes-photos-app-tags-two-black-people-gorillas}.

\bibitem[{Gul(2019)}]{gul_2019}
Gul, E. 2019.
\newblock Is Artificial Intelligence the frontier solution to Global South's
  wicked development challenges?
\newblock
  \urlprefix\url{https://towardsdatascience.com/is-artificial-intelligence-the-frontier-solution-to-global-souths-wicked-development-challenges-4206221a3c78}.

\bibitem[{Hara et~al.(2018)Hara, Adams, Milland, Savage, Callison-Burch, and
  Bigham}]{hara18amt}
Hara, K.; Adams, A.; Milland, K.; Savage, S.; Callison-Burch, C.; and Bigham,
  J. 2018.
\newblock A Data-Driven Analysis of Workers' Earnings on Amazon Mechanical
  Turk.
\newblock \emph{ACM Conference on Human Factors in Computing Systems (CHI)} .

\bibitem[{Harris, Straker, and Pollock(2017)}]{harris2017socioeconomic}
Harris, C.; Straker, L.; and Pollock, C. 2017.
\newblock A socioeconomic related'digital divide'exists in how, not if, young
  people use computers.
\newblock \emph{PloS one} 12(3): e0175011.

\bibitem[{Heston and Zwetsloot(2020)}]{cset}
Heston, R.; and Zwetsloot, R. 2020.
\newblock Mapping U.S. Multinationals’ Global AI R\&D Activity.
\newblock
  \urlprefix\url{https://cset.georgetown.edu/research/mapping-u-s-multinationals-global-ai-rd-activity/}.

\bibitem[{Jarosz(2003)}]{jarosz_human_2003}
Jarosz, L. 2003.
\newblock A {Human} {Geographer}'s {Response} to {Guns}, {Germs}, and {Steel}:
  {The} {Case} of {Agrarian} {Development} and {Change} in {Madagascar}.
\newblock \emph{Antipode} 35(4): 823--828.
\newblock ISSN 1467-8330.
\newblock \doi{https://doi.org/10.1046/j.1467-8330.2003.00356.x}.
\newblock
  \urlprefix\url{http://onlinelibrary.wiley.com/doi/abs/10.1046/j.1467-8330.2003.00356.x}.
\newblock \_eprint:
  https://onlinelibrary.wiley.com/doi/pdf/10.1046/j.1467-8330.2003.00356.x.

\bibitem[{Jo and Gebru(2020)}]{jo20archives}
Jo, E.~S.; and Gebru, T. 2020.
\newblock Lessons from Archives: Strategies for Collecting Sociocultural Data
  in Machine Learning.
\newblock \emph{ACM Conference on Fairness, Accountability, Transparency
  (FAccT)} .

\bibitem[{Jobin, Ienca, and Vayena(2019)}]{jobin_global_2019}
Jobin, A.; Ienca, M.; and Vayena, E. 2019.
\newblock The global landscape of {AI} ethics guidelines.
\newblock \emph{Nature Machine Intelligence} 1(9): 389--399.
\newblock ISSN 2522-5839.
\newblock \doi{10.1038/s42256-019-0088-2}.
\newblock \urlprefix\url{http://www.nature.com/articles/s42256-019-0088-2}.

\bibitem[{Kaye(2019)}]{kaye19dataannot}
Kaye, K. 2019.
\newblock These companies claim to provide “fair-trade” data work. Do they?
\newblock
  \urlprefix\url{https://www.technologyreview.com/2019/08/07/133845/cloudfactory-ddd-samasource-imerit-impact-sourcing-companies-for-data-annotation/}.

\bibitem[{Kelly(2020)}]{kelly_andrew_2020}
Kelly, M. 2020.
\newblock Andrew {Yang} is pushing {Big} {Tech} to pay users for data.
\newblock
  \urlprefix\url{https://www.theverge.com/2020/6/22/21298919/andrew-yang-big-tech-data-dividend-project-facebook-google-ubi}.

\bibitem[{Krasin et~al.(2017)Krasin, Duerig, Alldrin, Ferrari, Abu-El-Haija,
  Kuznetsova, Rom, Uijlings, Popov, Veit, Belongie, Gomes, Gupta, Sun, Chechik,
  Cai, Feng, Narayanan, and Murphy}]{openimages}
Krasin, I.; Duerig, T.; Alldrin, N.; Ferrari, V.; Abu-El-Haija, S.; Kuznetsova,
  A.; Rom, H.; Uijlings, J.; Popov, S.; Veit, A.; Belongie, S.; Gomes, V.;
  Gupta, A.; Sun, C.; Chechik, G.; Cai, D.; Feng, Z.; Narayanan, D.; and
  Murphy, K. 2017.
\newblock OpenImages: A public dataset for large-scale multi-label and
  multi-class image classification.
\newblock \emph{Dataset available from https://github.com/openimages} .

\bibitem[{Lee(2018)}]{lee2018big}
Lee, D. 2018.
\newblock Why Big Tech pays poor Kenyans to teach self-driving cars.
\newblock \emph{BBC News} .

\bibitem[{Lee et~al.(2019)Lee, Kusbit, Kahng, Kim, Yuan, Chan, See,
  Noothigattu, Lee, Psomas, and Procaccia}]{lee_webuildai_2019}
Lee, M.~K.; Kusbit, D.; Kahng, A.; Kim, J.~T.; Yuan, X.; Chan, A.; See, D.;
  Noothigattu, R.; Lee, S.; Psomas, A.; and Procaccia, A.~D. 2019.
\newblock {WeBuildAI}: {Participatory} {Framework} for {Algorithmic}
  {Governance}.
\newblock \emph{Proceedings of the ACM on Human-Computer Interaction} 3(CSCW):
  181:1--181:35.
\newblock \doi{10.1145/3359283}.
\newblock \urlprefix\url{http://doi.org/10.1145/3359283}.

\bibitem[{Lin(2011)}]{lin_flying_2011}
Lin, J.~Y. 2011.
\newblock From {Flying} {Geese} to {Leading} {Dragons} : {New} {Opportunities}
  and {Strategies} for {Structural} {Transformation} in {Developing}
  {Countries}.
\newblock Technical Report WPS 5702, World Bank.

\bibitem[{Masakhane(2021)}]{noauthor_masakhane_nodate}
Masakhane. 2021.
\newblock Masakhane: A grassroots NLP community for Africa, by Africans.
\newblock \urlprefix\url{https://www.masakhane.io/}.

\bibitem[{Mbayo(2020)}]{mbayo_2020}
Mbayo, H. 2020.
\newblock Data and Power: AI and Development in the Global South.
\newblock
  \urlprefix\url{https://www.oxfordinsights.com/insights/2020/10/2/data-and-power-ai-and-development-in-the-global-south}.

\bibitem[{Mendes, Bertella, and Teixeira(2014)}]{mendes_industrialization_2014}
Mendes, A. P.~F.; Bertella, M.~A.; and Teixeira, R. F. A.~P. 2014.
\newblock Industrialization in {Sub}-{Saharan} {Africa} and import substitution
  policy.
\newblock \emph{Revista de Economia Política} 34(1): 120--138.
\newblock ISSN 0101-3157.
\newblock \doi{10.1590/S0101-31572014000100008}.
\newblock
  \urlprefix\url{http://www.scielo.br/scielo.php?script=sci_arttext&pid=S0101-31572014000100008&lng=en&tlng=en}.

\bibitem[{Murgia(2019)}]{murgia2019ai}
Murgia, M. 2019.
\newblock AI’s new workforce: the data-labelling industry spreads globally.
\newblock \emph{Financial Times} .

\bibitem[{Nature(1915)}]{natureIRL}
Nature. 1915.
\newblock Industrial Research Laboratories.
\newblock \urlprefix\url{https://doi.org/10.1038/096419a0}.

\bibitem[{Nzekwe(2019)}]{nzekwe2019}
Nzekwe, H. 2019.
\newblock Africans Are Paying More For Internet Than Any Other Part Of The
  World – Here’s Why.
\newblock
  \urlprefix\url{https://weetracker.com/2019/10/22/africans-pay-more-for-internet-than-other-regions/}.

\bibitem[{Obermeyer et~al.(2019)Obermeyer, Powers, Vogeli, and
  Mullainathan}]{obermeyer2019dissecting}
Obermeyer, Z.; Powers, B.; Vogeli, C.; and Mullainathan, S. 2019.
\newblock Dissecting racial bias in an algorithm used to manage the health of
  populations.
\newblock \emph{Science} 366(6464): 447--453.

\bibitem[{O'Neill(2020)}]{jacki}
O'Neill, J. 2020.
\newblock Jacki O'Neill | LinkedIn.
\newblock \urlprefix\url{https://www.linkedin.com/in/jacki-o-neill-5605534/}.

\bibitem[{Paullada et~al.(2020)Paullada, Raji, Bender, Denton, and
  Hanna}]{paullada20data}
Paullada, A.; Raji, I.~D.; Bender, E.~M.; Denton, E.; and Hanna, A. 2020.
\newblock Data and its (dis)contents: A survey of dataset development and use
  in machine learning research.
\newblock \emph{arXiv:2012.05345} .

\bibitem[{Poskett(2013)}]{poskett_django_2013}
Poskett, J. 2013.
\newblock Django {Unchained} and the racist science of phrenology {\textbar}
  {James} {Poskett}.
\newblock \emph{The Guardian} ISSN 0261-3077.
\newblock
  \urlprefix\url{https://www.theguardian.com/science/blog/2013/feb/05/django-unchained-racist-science-phrenology}.

\bibitem[{Prabhu and Birhane(2020)}]{Prabhu20}
Prabhu, V.~U.; and Birhane, A. 2020.
\newblock Large image datasets: A pyrrhic win for computer vision?
\newblock \emph{arXiv:2006.16923} .

\bibitem[{Ranger(2001)}]{ranger_colonialism_2001}
Ranger, T. 2001.
\newblock Colonialism, {Consciousness} and the {Camera}.
\newblock \emph{Past \& Present} 203--215.
\newblock ISSN 0031-2746.
\newblock \urlprefix\url{http://www.jstor.org/stable/3600818}.
\newblock Publisher: [Oxford University Press, The Past and Present Society].

\bibitem[{Research(2020)}]{jackimsr}
Research, M. 2020.
\newblock Jacki O'Neill at Microsoft Research.
\newblock
  \urlprefix\url{https://www.microsoft.com/en-us/research/people/jaoneil/}.

\bibitem[{Rodney(1972)}]{rodney_how_1972}
Rodney, W. 1972.
\newblock \emph{How {Europe} underdeveloped {Africa}}.
\newblock London :: Bogle L'Ouverture Publications.
\newblock ISBN 978-0-9501546-4-0.

\bibitem[{Rogstadius et~al.(2011)Rogstadius, Kostakos, Kittur, Smus, Laredo,
  and Vukovic}]{rogstadius2011motivation}
Rogstadius, J.; Kostakos, V.; Kittur, A.; Smus, B.; Laredo, J.; and Vukovic, M.
  2011.
\newblock An Assessment of Intrinsic and Extrinsic Motivation on Task
  Performance in Crowdsourcing Markets.
\newblock \emph{Proceedings of the Fifth International Conference on Weblogs
  and Social Media} .

\bibitem[{Russakovsky et~al.(2015)Russakovsky, Deng, Su, Krause, Satheesh, Ma,
  Huang, Karpathy, Khosla, Bernstein, Berg, and Fei-Fei}]{imagenet}
Russakovsky, O.; Deng, J.; Su, H.; Krause, J.; Satheesh, S.; Ma, S.; Huang, Z.;
  Karpathy, A.; Khosla, A.; Bernstein, M.; Berg, A.~C.; and Fei-Fei, L. 2015.
\newblock {ImageNet Large Scale Visual Recognition Challenge}.
\newblock \emph{International Journal of Computer Vision (IJCV)} 115(3):
  211--252.
\newblock \doi{10.1007/s11263-015-0816-y}.

\bibitem[{Samasource(2021)}]{samasourceCauseIQ}
Samasource. 2021.
\newblock Samasource.
\newblock
  \urlprefix\url{https://www.causeiq.com/organizations/samasource,262547062/}.

\bibitem[{Sambasivan et~al.(2020)Sambasivan, Arnesen, Hutchinson, and
  Prabhakaran}]{sambasivan_non-portability_2020}
Sambasivan, N.; Arnesen, E.; Hutchinson, B.; and Prabhakaran, V. 2020.
\newblock Non-portability of {Algorithmic} {Fairness} in {India}.
\newblock \emph{arXiv:2012.03659 [cs]}
  \urlprefix\url{http://arxiv.org/abs/2012.03659}.
\newblock ArXiv: 2012.03659.

\bibitem[{Semuels(2018)}]{semuels2018internet}
Semuels, A. 2018.
\newblock The Internet Is Enabling a New Kind of Poorly Paid Hell.
\newblock \emph{The Atlantic}
  \url{https://www.theatlantic.com/business/archive/2018/01/amazon-mechanical-turk/551192/}.

\bibitem[{Shankar et~al.(2017)Shankar, Halpern, Breck, Atwood, Wilson, and
  Sculley}]{shankar17}
Shankar, S.; Halpern, Y.; Breck, E.; Atwood, J.; Wilson, J.; and Sculley, D.
  2017.
\newblock No Classification without Representation: Assessing Geodiversity
  Issues in Open DataSets for the Developing World.
\newblock \emph{NeurIPS workshop: Machine Learning for the Developing World} .

\bibitem[{Silva(2021)}]{BlackInAI}
Silva, M. 2021.
\newblock \urlprefix\url{https://blackinai.github.io/#/about}.

\bibitem[{Sloane et~al.(2020)Sloane, Moss, Awomolo, and
  Forlano}]{sloane_participation_2020}
Sloane, M.; Moss, E.; Awomolo, O.; and Forlano, L. 2020.
\newblock Participation is not a {Design} {Fix} for {Machine} {Learning}.
\newblock \emph{arXiv:2007.02423 [cs]}
  \urlprefix\url{http://arxiv.org/abs/2007.02423}.
\newblock ArXiv: 2007.02423.

\bibitem[{Synced(2019)}]{synced}
Synced. 2019.
\newblock Data Annotation: The Billion Dollar Business Behind AI Breakthroughs.
\newblock
  \urlprefix\url{https://medium.com/syncedreview/data-annotation-the-billion-dollar-business-behind-ai-breakthroughs-d929b0a50d23}.

\bibitem[{Thompson(2016)}]{thompson_otherness_2016}
Thompson, A. 2016.
\newblock Otherness and the {Fetishization} of {Subject}.
\newblock
  \urlprefix\url{https://petapixel.com/2016/11/16/otherness-fetishization-subject/}.

\bibitem[{Wang, Narayanan, and Russakovsky(2020)}]{wang20revise}
Wang, A.; Narayanan, A.; and Russakovsky, O. 2020.
\newblock {REVISE}: A Tool for Measuring and Mitigating Bias in Visual
  Datasets.
\newblock \emph{European Conference on Computer Vision (ECCV)} .

\end{thebibliography}

% \section{Acknowledgments}

\end{document}